\documentclass[conference]{IEEEtran}
\IEEEoverridecommandlockouts
\usepackage{cite}
\usepackage{amsmath,amssymb,amsfonts}
\usepackage{algorithm} 
\usepackage{algorithmic}
\usepackage{graphicx}
\usepackage{textcomp}
\usepackage{xcolor}
\usepackage[a4paper, total={184mm,239mm}]{geometry}
\def\BibTeX{{\rm B\kern-.05em{\sc i\kern-.025em b}\kern-.08em
		T\kern-.1667em\lower.7ex\hbox{E}\kern-.125emX}}
\usepackage{fancyhdr} 
\renewcommand{\headrulewidth}{0pt} 

\begin{document}

\fancypagestyle{plain}{ 
	\fancyhead{} 
	\fancyhead[L]{To appear in the proc. of the 35th Intern. IEEE Workshop on Rapid System Prototyping, 3rd Oct. 2024, Raleigh, NC, USA} 
	\renewcommand{\headrulewidth}{0pt} 
}

\title{Non-interfering On-line and In-field SoC Testing}

\author{\IEEEauthorblockN{Tobias Strauch}
	R\&D, EDAptix e.K.\\
	Munich, Germany\\
    Email: tobias@edaptix.com}

\maketitle
\maketitle \thispagestyle{plain}

\begin{abstract}
With increasing aging problems of advanced technologies, in-field testing becomes an inevitable challenge, on top of the already demanding requirements, such as the ISO26262 for automotive safety. SOCs used in space, automotive or military applications in particular are worst affected as the in-field failures in these applications could even be life threatening. We focus on on-line and in-field testing for Single Event Upsets (SEU, caused by a single ionizing particle) and aging defects (such as delay variation and stuck-at faults) which may appear during normal operation of the device. Interrupting normal operations for aging defects testing is a major challenge for the OS. Additionally, checkpointing with rollback-recovery can be costly and mission critical data can be lost in case of an SEU event. We eliminate many of these problems with our non-interfering in-field testing and recovery solution.

We apply a hardware performance improvement technique called System Hyper Pipelining (SHP), which combines well-known context switching (Barrel CPU) and C-slow retiming techniques. The SoC is enhanced with an SEU detection and ultra-fast recovery mechanism. We also use an RTL ATPG framework that enables the generation of software-based self-tests to achieve 100\% coverage of all testable stuck-at-faults. The paper finishes with very promising performance-per-area and test-cycles-per-net results. We argue that our robust system architecture and EDA solution, designed and developed primarily for in-field testing of SoCs, can also be used for production and on-line testing as well as other applications.

\end{abstract}


\section{Introduction}
Chips used in aerospace, automotive, and military applications are subject to in-field failures that can be extremely mission costly or even life-threatening.

Cosmic ray phenomena such as solar particle events cause high radiant flux that lasts for hours to days, increasing the likelihood of single-event upsets (SEUs) by several orders of magnitude. With the advent of nanoscale (high-)performance computing, soft errors that impact the reliability of modern electronic systems even at ground level have become one of the most challenging issues for the semiconductor industry.

All parts of a design can be affected, including neural networks, where Failure In Time (FIT) rates can exceed safety standards, e.g. ISO 26262 for the automotive industry, by orders of magnitude, as shown in \cite{AI_SEU}.

There are also device defects that can occur during in-field operation of the device and are mainly due to latent faults that may not be obvious or readily detectable during production or on-line testing but may develop over time under real-time applications in the field due to environmental conditions.

The industry is responding to these challenges with standardizations such as ISO 26262. These new requirements must coexist with existing applications and testing must be carefully scheduled to avoid impacting the applications on the device. Efficient scheduling for on-line and in-field testing can be a major challenge for the operating system as \cite{SCHED_ROB} and \cite{SCHED_DYN} clearly demonstrate. Additionally, checkpointing with rollback-recovery can be costly (power, timing, …) \cite{REC_MC} and mission critical data can be lost in case of an SEU event when a system rollback must be initiated.

In this paper, we introduce a robust SoC architecture and EDA software solution to cope with the aforementioned challenges. The main goal is to continuously test for SEU and aging faults during on-line testing and in-field operation without interfering with the normal operation as well as to recover from an SEU detection very efficiently. 

In order to provide a self-contained work, we start the paper with a list of short introductions to the respective techniques on which this work is based, such as 
\begin{itemize}
	\item an interleaved multithreading technique (Section II),
	\item functional redundancy and failure recovery (Section III),
	\item aging-related failure detection (Section IV) and
	\item a gate inherent fault based RTL ATPG (Section IV).
\end{itemize}
Our work is introduced in Section VI and compared to related work in Section VII, before results are presented in Section VII.

\section{Barrel CPU and C-slow Retiming}
Fig. \ref{fig_sketches}a shows the basic structure of a sequential circuit with its combinational logic (CL) and original design registers (DR). Clock, in- and outputs are not shown for the sake of simplicity. The sequential circuit processes a single thread T(0) running at what we define here as macro-cycle speed. 

\subsubsection{Barrel CPU}

A barrel processor is a CPU that switches between threads of execution every cycle. The design technique is also known as "interleaved" or "fine-grained" temporal multithreading. A modern example of a barrel RISC-V CPU is shown in \cite{barvin}. 

Fig. \ref{fig_sketches}b gives an abstract view of a design based on the barrel technique. The DRs are now replaced by memories (Mem) and the design is extended by a thread controller (TC). D is the number of threads the memory can hold (memory depth). The executed thread can now be freely selected within D threads (read pointer) and saved at the corresponding address (write pointer) using the thread index. The individual threads still run at macro-cycle speed.

\subsubsection{C-slow retiming (CSR)}
The C-slow retiming (CSR) technique provides C copies of a given design by inserting an equal number of registers into each combinatorial path and therefore reusing the logic in a time sliced fashion \cite{csr}.  

\begin{figure}[!t]
	\centering
	\includegraphics[width=3.3in,height=7cm]{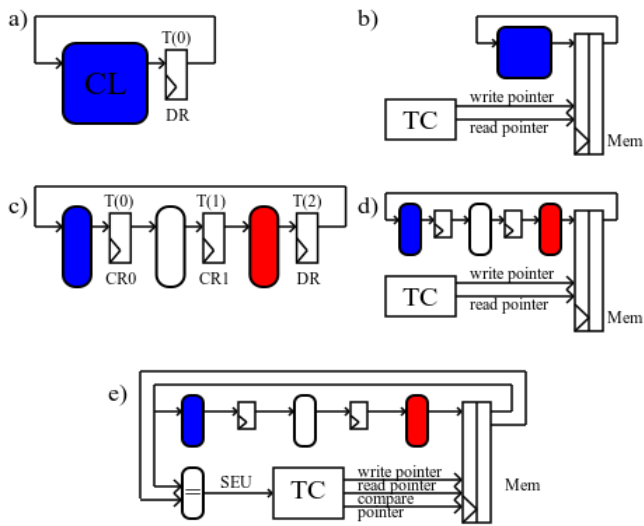}
	\caption{a) Simplified single clock design. b) Applying barrel technique c) Applying C-slow retiming d) Applying System-Hyper-Pipeling (SHP) e) SEU detection and recovery based on C-slow retiming and applyed on SHP}
	\label{fig_sketches}
\end{figure}

Fig. \ref{fig_sketches}c outlines the CSR technique. The original logic is sliced into C (here C=3) sections, and each original path now has C-1 additional registers running at micro-cycle speed. This results in C functional independent design copies T(0, ..., C-1) which use the logic in a time sliced fashion. Each thread has its own thread index. For each design copy it now takes C micro-cycles to achieve the same result as in one cycle of the original design (macro-cycle). The implemented register sets are called ``CSR Registers`` (CR). 

\subsubsection{System Hyper Pipelining (SHP)}

System hyper-pipelining (SHP) is a technique introduced in \cite{SHP} that combines the barrel and C-slow retiming techniques mentioned above. Fig. \ref{fig_sketches}d shows the modifications towards an SHP-ed design, which can run any number of threads (T \textless= D) in any possible interleaved order.

\subsubsection{Thread mixing}

When applying SHP on an SoC, the number of individual threads (D, barrel technique) as well as the number of interleaved executed design copies C for individual subblocks can vary. For example, the less timing critical Ethernet design does not need to use C-slow retiming to achieve the required performance and therefore only the barrel technique needs to be applied. Accelerators, on the other hand, are usually time-critical and only the C-slow retiming technique might be relevant. The CPU is based on SHP to achieve the best possible performance-per-area trade-off. This approach allows for an optimal thread mixing and best serves our purpose while providing an optimal performance-per-area trade-off.

\subsubsection{Load balancing}
\begin{figure}[!t]
	\centering
	\includegraphics[width=3.5in,height=4cm]{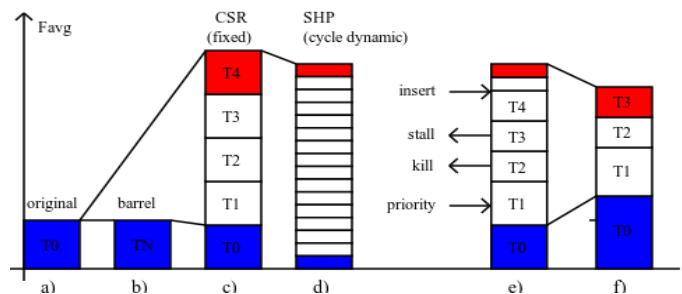}
	\caption{Average thread performance (Favg) of different scenarios running a) Original design, b) Design with barrel, c) C-slow retiming and d-f) SHP technique .}
	\label{fig_histo}
\end{figure}

\setlength{\arraycolsep}{5pt}
Fig. \ref{fig_histo} shows the advantages of the aforementioned techniques compared to the original design. The x-axis of the histogram shows different scenarios/solutions, the y-axis the system performance. Assuming a thread (T0) on the original CPU runs at e.g. 80MHz on an FPGA (Fig. \ref{fig_histo}a). 

The barrel CPU version allows context switching between multiple threads, but does not improve CPU performance as such (Fig. \ref{fig_histo}b) as it still runs at macro-cycle speed.

It can be seen how CSR improves the system performance of the original system implementation (Fig. \ref{fig_histo}c). System performance is no longer necessarily limited by the critical path of the original design or external memory access, but rather, for example, by the switching limit of the FPGA (e.g. 600 MHz). The design runs at micro-cycle speed. When using CSR, all threads run at the same speed and load balancing is not possible.

\indent For executing multiple programs on multiple CPUs (symmetrical multi-processing), SHP allows a more efficient usage of the system resources (Fig. \ref{fig_histo}d to \ref{fig_histo}f). It adds the possibility to distribute the system performance over a minimum (C, Fig. \ref{fig_histo}c) and a maximum set of threads (D, Fig. \ref{fig_histo}d), whereas any solution in-between can be realized. Fig. \ref{fig_histo}e) shows a random example. This load balancing is handled by a TC and can be dynamically modified during runtime. Fig. \ref{fig_histo}f refers to more advanced SHP techniques as shown in \cite{SHP}, where more system performance is given to specific threads.


\section{Redundancy and failure recovery}
\setlength{\arraycolsep}{5pt}

So far we have briefly described well-known digital design concepts such as a barrel CPU and C-slow retiming as well as their combined application (SHP). These concepts can now be extended to detect and to recover from SEUs.

An SEU is a change of state caused for example by a single ionizing particle (ion, electron, photon...) hitting a sensitive node in a design. The change of state is a result of the free charge created by ionization in or near an important node of a logic element (e.g. register). The failure in device output or operation caused by the strike is called SEU.

The main techniques to detect an SEU are either based on spatial redundancy or temporal redundancy. Spatial redundancy is based on the replication of n-times the original module building n+1 identical redundant modules, where outputs are merged into a majority voter. Time redundancy is based on capturing the states multiple times to vote out a transient fault. The values are shifted by a delay. The idea is to be able to capture a majority of upset free values to be able to mask the fault. We define the level of redundancy as R.

Some approaches based on time redundancy use interleaved-multi-threading to detect and to recover from such an SEU. In a recent publication \cite{BAR_REC} the aforementioned barrel technique is applied on a RISC-V processor on selected CPU elements, such as program counter, register file, etc.. Identical threads are executed, and the results are compared. If a mismatch is detected, a recovery mechanism restores the system, using an auxiliary thread as reference.

The same basic idea is shown in \cite{CSR_REC}, based on designs which use the aforementioned C-slow retiming technique. C-slow retiming inserts the same number of registers into each path to use the logic in a time-sliced fashion. It is demonstrated, how to enhance such a design with an SEU detection logic, how identical threads can be executed on such a design and how a design can recover from an SEU fault within a limited number of cycles. Fig. \ref{fig_sketches}e shows the basic concept of our work (inspired by \cite{CSR_REC}) with the extension that the design registers are replaced by memories and the TC controls the recovery sequence after an SEU detection.


\section{Detection of aging-related failures}

\subsubsection{Using timing critical path measurement} 
Aging related Bias Temperature Instability (BTI) and Hot Carrier Injection (HCI) faults affect the delay of individual cells and the overall path timing, as shown in \cite{aging_path1} and \cite{aging_path2}. An experiment using ring oscillators to demonstrate these aging effects on an FPGA is shown in \cite{aging_ring_fpga}.

Aging introduced faults can be modeled on RTL \cite{aging_path_rtl_model} and can already be considered during logic synthesis \cite{aging_timing_violation_path} by shortening critical paths or during the design, place and route steps when FPGAa are used \cite{fpga_aging_rates}.

Testing delay faults in functional mode is shown in \cite{aging_functional_mode_delay}. The selection of timing critical paths for this purpose can be achieved by assertion guided SBST \cite{selection_assertions} during RTL verification with the help of statistical timing models (such as \cite{selection_critical_path}) or on gate level using gate level timing information \cite{selection_timing_violation}.

The hardware (HW) can be enhancement for critical delay measurement as shown in \cite{hw_delay_meassurement} or through robust and in-situ self-testing techniques outlined in \cite {hw_insitu_critical_path_monitoring}.

Since traditional test methods interfere with normal operation, the issue of scheduling test tasks becomes very critical. Calculating periodic testing in embedded processors is discussed in \cite{scheduling_periodic_modeling} and at-speed tests using functional tests for delay faults is evaualted in  \cite{scheduling_delay_faults}. Also power related issues during on-line and in-field testing \cite{scheduling_online_power_aware} as well as OS related challenges \cite{scheduling online_task} must be considered.

\subsubsection{Using transistor activation and propagation}

Aging-related delay variations can also affect hold time, pulse width, and other timing requirements that certainly cannot be continuously measured. Another aging-related problem is the time-dependent dielectric breakdown (TDDB) of transistors \cite{dielectric_breakdown}, which ultimately leads to a fatal failure of the transistor. We argue that activation and comparison with a reference value (after propagation) of all testable signals (also known as stuck-at testing), helps to identify aging-related transistor TDDB failures that cannot be detected by critical path measurement.

\section{Gate Inherent Fault (based RTL ATPG)}

The term “RTL ATPG” defines the methodology for generating stuck-at faults (SAF) test patterns based on RTL design descriptions. This can be done using dedicated functional tests which are executed on the device. An overview of software based self-tests (SBST) is given in \cite{overview_sbst}. A promising Gate Inherent Fault (GIF) RTL ATPG model was presented in \cite{gif}. 

The GIFs are extracted from each complex RTL primitive (multiplier, adder, shifter, case-statement, etc.) of the RTL source code individually. They are related to the internal logic paths of a complex gate. They are not related to any net/signal or gate in the gate level design. It is observed that when all GIFs on RTL are covered (100\%) and the same stimulus is applied on gate level, then all testable gate level SAFs of the netlist are covered (100\%) as well. The GIF model is therefore synthesis independent.

The GIF model can be applied on any alternative language construct (multiplier, etc.) or any combination of language constructs as well. The key point is, that the GIF model is related to internal paths of complex gates and not to signals in the given RTL design nor to nets in its gate level representation.




\subsection{The GIF-GO model definition}
\indent Under the proposed GIF gate output (GIF-GO) model, a GIF is described by a quadruple (gi, go, i, $\alpha$) where gi is a gate input, go is a gate output, i is an index and $\alpha$ $\in$ \{0,1\}. The fault (gi, go, i, $\alpha$) is detected by a test t that satisfies the following conditions:

\indent \space 1. The test t detects the path fault gi to go with index i\\ 
\indent \space \space \space \space (gi-go-i).\\
\indent \space 2. The fault free value of \textbf{gate output go} under t is $\alpha$.\\
\indent \space 3. In the presence of the fault gi-go-i, the output value\\ 
\indent \space \space \space \space go = !$\alpha$.

In other words, t propagates the effects of a gi-go path fault with index i to the gate output go. The output's value is $\alpha$ in the fault free circuit and !$\alpha$ in the presence of the fault. 

An alternative view on this argues that the functionality of each (complex) gate can be defined by a Karnaugh map. The GIF model now states, that each ‘1’ and ’0’ entry of the gate’s Karnaugh map must be sensitized and propagated to primary outputs.

\subsection{Logic duplication}
\indent An important element of RTL synthesis is logic duplication. Duplicated logic can generate net faults which are not detected when a test set is used that is based on the GIF-GO model. Therefore the final RTL fault model needs to consider logic duplication. All outputs of a design are called primary outputs (PO). In case of a sequential netlist, register data inputs are considered as PO as well.

\subsection{The GIF-PO model definition}
\indent Under the proposed GIF-PO model, a GIF is described by a quintuple (gi, go, i, j, $\alpha$) where gi is a gate input, go is a gate output, i is an index, j is a primary output and $\alpha$ $\in$ \{0,1\}. The fault (gi, go, i, j, $\alpha$) is detected by a test t that satisfies the following conditions:\\
\indent \space 1. The test t detects the path fault gi to go with index i\\ 
\indent \space \space \space \space (gi-go-i).\\
\indent \space 2. The fault free value of \textbf{primary output j} under t is $\alpha$.\\
\indent \space 3. In the presence of the fault gi-go-i, the primary output\\ 
\indent \space \space \space \space value j = !$\alpha$.\\
\indent In other words, t propagates the effects of a gi-go path fault with index i to the primary output j. The primary output's value is $\alpha$ in the fault free circuit and !$\alpha$ in the presence of the fault.


\section{Our work}

Efficient scheduling for on-line and in-field testing can be a major challenge for the operating system as \cite{SCHED_ROB} and \cite{SCHED_DYN} clearly demonstrate. Additionally, checkpointing with rollback-recovery can be costly (power, timing, …) \cite{REC_MC} and mission critical data can be lost in case of an SEU event when a system rollback must be initiated.

The unique contribution of our work is that we demonstrate how an interleaved multithreaded SHP architecture can be utilized for non-interfering on-line and in-field testing. Without interrupting normal operations, we demonstrate
\begin{itemize}
	\item how to detect and recover from SEU faults,
	\item how to detect faults generating functional mismatches and
	\item how to detect delay faults caused by aging.
\end{itemize}

As far as the authors are aware, this simultaneous approach has not been proposed before. The following steps are executed:

\subsection{Our work: hardware related}

\subsubsection{SoC specification and preparation}
For each element of the SoC, such as the CPU, communication, and acceleration peripherals, etc., we individually specify the parameters C and D, where C refers to the number of design copies we achieve by applying C-slow retiming and D refers to the number of threads we want to store (barrel technique, memory depth).

\subsubsection{Applying barrel technique (manually)}
We then manually improve the design by replacing registers with a set of registers (or memory bits) and by adding the appropriate read and write logic to the design for individual thread execution.

\subsubsection{Applying C-slow retiming technique (automatically)}
The design is automatically improved by incorporating the C-slow retiming technique. This timing driven automatic register insertion technique is performed on RTL as presented in \cite{CSR_TIMING}.

\subsubsection{Inserting SEU detection and recovery logic} 
The design is further manually optimized to support the SEU detection and recovery mechanism similar to the concept shown in \cite{CSR_REC}.
 
Memory read port:
Based on the modification to support the barrel technique, design states of individual threads are stored in memories or small register sets, depending on the number of maximum threads (memory depth, D). Redundant threads may be stored in locations as far apart as possible. Additionally, a 2nd read port is added to the memories. 
 
Thread controller: 
The TC drives the write port to store a design state (or not). It also controls the read ports to a) start execution of a thread cycle and b) to compare its state with a second (redundant) thread at the beginning of a cycle execution (see Fig. \ref{fig_sketches}e).

Comparison logic:
Additional comparison logic detects mismatches between the two selected threads. This logic can be pipelined similar to the scheme used for C-slow retiming.

Algorithm:
The algorithm for detecting and recovering from an SEU is based on the concept that redundant thread cycles are only completed (stored) when all threads start with identical state values. All threads start from individual memory locations. These starting state values are then compared, while the threads are propagated through the C-slow retimed logic. 

If no mismatch is discovered, the resulting state is written R times and normal operation continues. If a mismatch is detected, no state is overwritten and the cycle is repeated. The TC recognizes the results of the majority voting and replaces the start conditions of faulty threads with one of the correct threads. 

Enhancements:
If the SEU detection period takes longer than the execution, then the executed threads can be stored in alternating memory locations to avoid overwriting valid thread states. Another improvement is to replace the additional read port with a pipelined state capture register and to update the associated comparison logic accordingly. The TC then ensures that two consecutive threads can be compared. These registers can be the same registers inserted for the C-slow retiming technique.

\subsubsection{HW adaption for RTL ATPG}
In order to continuously test for SAFs certain HW related optimizations are required. These can be features like loop-back logic or an overwriting mechanism to set a counter into a defined state by software. This will become clearer in the next section.

\subsection{Our work: EDA software related}
\label{sec_rtl_atpg}

In this section, we present a framework based on an advanced RTL simulator and a coverage database viewer. The goal is to generate functional tests that can be run on the device during on-line testing or in-field operation and collect the maximum number of GIFs when executed. The RTL simulator recognizes all relevant GIFs of the source code and passes their coverage throughout the logic during functional simulation. Sequential functional tests typically stimulate and propagate GIFs over many execution cycles until they can be observed at relevant registers by the application running on the device. In other words, the test result should be different in the presence of a fault compared to the fault free behavior.

On the SHP-based HW, GIF-related threads do not interfere with normal operation and can be scheduled to run in parallel. It can be beneficial to test more safety-critical logic, such as control logic, more frequently than less critical sections, such as an FPU for instance.

Due to the complexity of the GIF test pattern generation on RTL, it is almost imperative to split the overall task into multiple test sets, most likely related to individual sub-designs. For each test, the GIF coverage characteristic is stored in a database and the results of a single test or multiple test runs can be analyzed using a database viewer. When a GIF cannot be covered, it is usually an indicator of redundant logic. 

\begin{figure}[!t]
	\centering
	\includegraphics[width=3.3in,height=9cm]{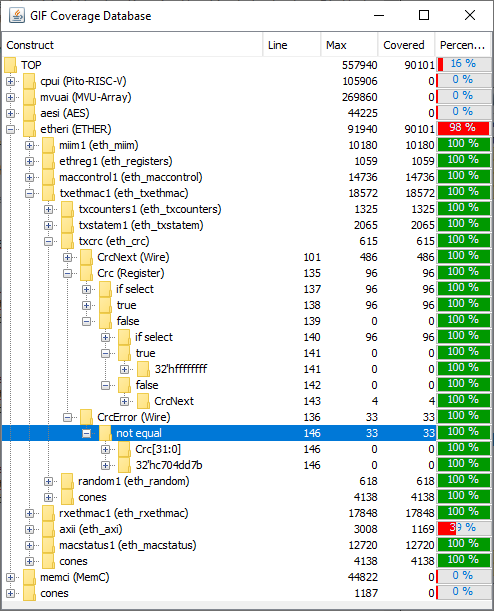}
	\caption{Snapshot of the coverage viewer GUI.}
	\label{fig2}
\end{figure}

An example of a database viewer is given in Fig. \ref{fig2}. The SoC design contains a CPU, an SDRAM controller as well as some peripherals such as an Ethernet core (here shown partly unfolded). All testcases related to this core are selected and their accumulated GIF coverage is displayed. It can be seen that some GIFs are covered (e.g. if-then-else or not-equal construct) and that the relative coverage on the Ethernet core itself reaches 98\%. The other cores have low GIF coverage because only Ethernet core related testcases are merged in this example.


Hard to cover faults are the reason why the test pattern generation process is usually accompanied by HW adjustment efforts. This includes the possibility of setting counter registers via SW. Also, including loopback functionality in communication peripherals is quite common for GIF testing purposes if it is not already present in the design. 

\section{Related work}
In \cite{BAR_REC}, a SEU detection and recovery mechanism is proposed based on the concept we referred to as the barrel technique in Section II. Our work follows the concept outlined in \cite{CSR_REC}, which uses C-slow retiming for interleaved multithreading. 

In \cite{BAR_REC}, redundant threads are executed and once a mismatch is detected, an auxiliary thread is used for recovery, which in turn may be subject to an SEU fault. In our work, no fixed auxiliary thread is used. It follows the rule that the states of all redundant threads are only overwritten when their start conditions (register values) are identical. Once a mismatch is detected, the thread controller replaces the failing thread with one of the remaining correct threads (not necessarily a single very specific thread) and initiates an ultra-fast recovery mechanism.

Approaches solely based on the barrel technique \cite{BAR_REC} reduce the system performance with each additional redundant thread due to insufficient logic sharing. In contrast, the advantage of the C-slow retiming approach is that there is only a small degradation in the maximal thread performance (due to register insertion) when running C threads on the system, while dramatically increasing the performance-per-area factor at the same time \cite{CSR_REC}.

It is not clear to the authors how the comparison logic proposed in \cite{BAR_REC} can be fast enough to detect SEU faults at the end of a single cycle. No performance results are presented in \cite{BAR_REC}. In our work, the SEU detection logic compares register values at the start of a cycle and the comparison logic can be pipelined following the C-slow retiming technique (Fig. \ref{fig_sketches}e). In an extended version, threads can be continuously stored in an alternating register bank to be used for normal operation but also to have a backup version after completion of the comparison task to be used by a simple rollback mechanism.



In \cite{sbst_atpg_processor_selftest} Riefert et al. demonstrate the use of SAT solvers within an RTL ATPG framework for SBST of in-field testing of a processor. The flow still depends on gate level faults and repetitive gate level fault simulation steps, which makes its usage for large SoCs questionable. In the GIF model-based solution the test pattern are generated entirely on RTL to generate test pattern for in-field execution with 100\% coverage of all testable SAFs on gate level.

Table \ref{tab_fc} shows the SAF coverage (SAFC) reported in the literature \cite{sbst_crypto, sbst_soc, sbst_vliw, sbst_mips, fc_embedded, fc_embedded2, fc_sbst, fc_sbst_abstr, fc_risc} for various IP blocks, which are used for SBST based SAF detection. Only one work reports 100\% SAFC. It is based on an AES example \cite{sbst_crypto}. With our demonstrated framework tests can be generated for 100\% coverage of all testable SAFs on the complete SoC for in-field testing guided by the database viewer in an interactive process.


Gao et al. \cite{TMOC} propose a Time-Multiplexed Online Checking (TMOC) scheme using embedded blocks for checker implementation, which enables various system parts to be checked dynamically during in-field operation in a time-multiplexed fashion. Also, a reliability analysis for optimal periodic testing of intermittent faults that minimizes the test cost was introduced by Kranitis et al. in \cite{PERIODIC}. It can be argued that with our interleaved and non-interfering solution task scheduling for on-line and in-field testing becomes less challenging. 




\section{Results}

\subsubsection{Design preparation, applying SHP}

\begin{table*}[!ht]
	\caption{SoC module performance-per-area results and test-cycle-per-net (TCPN) for FPGA and ASIC implementations.}
	\begin{center}
		\begin{tabular}{|c|c|c|c|c|c|c|c|c|c|c|c|c|c|c|c|c|c|}
			\hline
			\multicolumn{2}{|c|}{}
			&\multicolumn{3}{c|}{FPGA, original}
			&\multicolumn{5}{c|}{FPGA, optimized}
			&\multicolumn{3}{c|}{ASIC, original}
			&\multicolumn{5}{c|}{ASIC, optimized}\\
			\hline
			&GIF&LUT&Perf&PpA&LUT&Perf&PpA&PpAr&&Area&Perf&PpA&Area&Perf&PpA&PpAr&\\
			&[k]&[k]&[M&[MHz&[k]&[M&[MHz&[\%]&TCPN&[k&[M&[MHz/&[k&[M&[MHz/&[\%]&TCPN\\
            &&&Hz]&/k]&&Hz]&/k]&&&µm\textsuperscript{2}]&Hz]&kµm\textsuperscript{2}]&µm\textsuperscript{2}]&Hz]&kµm\textsuperscript{2}]&&\\
			\hline
			CPU & 106 & 10.4 & 250 & 23.9 & 13.0 & 675 & 51.8 & 2.16 & 0.97 &
			            65.3 & 155 & 2.38 & 117  & 385 & 3.27 & 1.37 & 2.31 \\
			\hline
			MVU & 270 & 31.7 & 250 & 7.87 & 41.8 & 598 & 14.3 & 1.81 & 0.87 &
			            186  & 150 & 0.80 & 297  & 337 & 1.13 & 1.41 & 1.61 \\
			\hline
			AES & 44.2 & 8.46 & 310 & 36.6 & 9.86 & 781 & 79.2 & 2.16 & 0.18 &
			             50.7 & 329 & 6.50 & 84.6 & 667 & 7.89 & 1.21 & 0.36 \\
			\hline
			Ether & 91.9 & 12.9 & 449 & 34.7 & 15.8 & 862 & 54.6  & 1.57 & 0.72 &
  			               85.5 & 341 & 3.99 & 169  & 739 & 4.36 & 1.09 & 1.97 \\
			\hline
			MemC & 44.8 & 6.75 & 296  & 43.8 & 7.98 & 795 & 99.6  & 2.27 & 0.32 &
			              45.6 & 247 & 5.43 & 92.8 & 556 & 5.99 & 1.10 & 1.05 \\
			\hline
		\end{tabular}
		\label{tab_results}
	\end{center}
\end{table*}



\begin{table}[htb]
	\caption{SAF coverage (SAFC) and test-cycles-per-net (TCPN) numbers.}
	\begin{center}
		\begin{tabular}{|c|c|c|c|c|c|c|c|c|c|c|c|}
			\hline
			&FPGA&ASIC&
			\cite{sbst_crypto}&
			\cite{sbst_soc}&
			\cite{sbst_vliw}&
			\cite{sbst_mips}\\
			\hline
			Source & SoC & SoC & AES & SoC per.& VLIW & MIPS \\
			\hline
			SAFC [\%]& 100  & 100 & 100 & 94.92 & 98.3 & 97.46 \\
			\hline
			TCPN & 0.61 & 1.46 & n.a. & n.a. & 0.024 & n.a. \\
			\hline
			\hline
			&\cite{fc_embedded}&
			\cite{fc_embedded2}&
			\cite{fc_sbst}&
			\cite{fc_sbst_abstr}&
			\cite{fc_risc}&\\
			\hline
			Source
			&\multicolumn{5}{c|}{Processor}&\\
			\hline
			SAFC [\%]& 92.3 & 92.7 & 90.03 & 93.74 & 92.2 &\\
			\hline
			TCPN & 9.19 & 0.13 & n.a. & 0.18 & 0.10 &\\
			\hline
		\end{tabular}
		\label{tab_fc}
	\end{center}
\end{table}

For our SoC reference design we use BARVINN \cite{barvin}, which is based on a barrel CPU (RISC-V) and a set of Matrix Vector Units (MVUs) optimized for AI algorithms. We also added a cryptographic (AES) and a communication (Ether) peripheral as well as an SDRAM memory controller (MemC). Table \ref{tab_results} shows the GIF number for each module.

We apply CSR (C=4) on the CPU and all peripherals. 33\% of the MVU designs can be removed as the remaining MVU blocks can now run in a time-sliced fashion. We also apply the barrel technique on the Ethernet core. 

We base our results on FPGA technology (AMD, Kintex) as FPGAs are used in space, automotive and military applications and use the term area synonymously with LUTs. Our reference design is also implemented with ASIC technology (Sky130). Here, the term area includes the standard cell area as well as the additional area resulting from the use of small memory cells required to support the SHP technology. For FPGA technology we chose D=16 and for ASIC technology D=8.

Table \ref{tab_results} shows the original and the optimized area as well as the respective performance. Based on that, the performance per area factor (PpA) is listed. The increase of the relative PpA number (PpAr) is also given for each design block.

In a multiple core lockstep configuration, the PpAr remains constant, whereas in our proposed system architecture, the PpAr improves significantly for both FPGAs and ASICs. The idea of the overall concept is to use this performance gain for non-interfering testing. The workload of the test application can then be adapted to suit on-line and in-field test requirements.

\subsubsection{SEU detection and Recovery}

All design blocks are capable of interleaved multi-threading supporting multiple identical subsequent threads. We chose to execute three redundant threads (R$=$3) and added a TC as well as the SEU detection logic mentioned above to the SoC. Since the number of redundant threads is less than the number of executed threads (R$<$C) and because the SEU detection logic is fast enough, no intermediate thread context storing is necessary.

Due to the SEU detection and recovery logic insertion, the FPGA SoC LUT count increased by 0.5\% in average and the average area increase for the ASIC is 0.6\%), which is not shown in Table \ref{tab_results}. Since our methodology is based on C-slow retiming, we expect the same advantages in power consumption compared to alternative approaches, as reported in \cite{CSR_REC}.

\subsubsection{Stuck-at detection}

We generated non-interfering testcases for each individual SoC block using the EDA software presented in Section \ref{sec_rtl_atpg}. We achieve 100\% SAF coverage of all testable faults on gate-level. The area impact of the HW enhancements is neglectable. 

Table \ref{tab_results} shows the test-cycles-per-net (TCPN) for each individual SoC block. SBST based SAF coverage is shown in Table \ref{tab_fc} for crypto-devices \cite{sbst_crypto} (100\% stuck-at-fault coverage, (SAFC)), SoC communication peripherals \cite{sbst_soc} (95\% SAFC) and processors \cite{sbst_vliw, sbst_mips, fc_embedded, fc_embedded2, fc_sbst, fc_sbst_abstr, fc_risc} (92.2\% - 98.2\% SAFC). We calculate an average TCPN for our FPGA implementation of 0.61 and 1.46 for the ASIC implementation respectively (listed in Table \ref{tab_fc}). Alternative work reported here with lower TCPN \cite{sbst_vliw, fc_embedded2, fc_sbst_abstr, fc_risc} do not reach 100\% SAFC and the design flow reported in \cite{sbst_vliw} is also highly optimized. 


\subsubsection{Fault injection simulation}

In our demonstrated methodology, SEU events and aging-induced errors that result in a functional sequential mismatch are detected through design state comparison. Any SAF caused by production or aging issues is detected through a comprehensive functional testing program, resulting in 100\% SAF coverage of all testable faults on gate level. Fault injection simulation does not provide any meaningful results in this context and is therefore not used.

\section{Summary}

To meet increasingly challenging safety requirements, SoCs must be designed to carry out in-field testing (ISO26262). Interrupting normal operations for aging defects testing is a major challenge for the OS. Additionally, checkpointing with rollback-recovery can be costly and mission critical data can be lost in case of an SEU event. To drastically reduce these problems, we use a robust system architecture based on an interleaved multi-threaded HW concept (system-hyper-pipelining, SHP), which combines the advantages of context switching (barrel technique) and C-slow retiming. We also enhance this structure by an SEU detection and fast recovery mechanism.

In this paper we concentrate on SEU detection and recovery as well as on delay measurement and SAF testing during normal in-field operation. The area overhead for inserting the SEU detection logic is extremely low and the recovery period is ultra-fast. Our proven RTL ATPG flow enables the generation of 100\% SAF tests of all testable faults and the area impact to perform these software-based tests is negligible. The tests do not interfere with normal operation and can be dynamically scheduled depending on the application’s workload and safety requirements.



\end{document}